Collisional disruption of highly porous targets in the strength regime: Effects of mixture

Yuichi Murakami[1], Akiko M. Nakamura[1], Koki Yokoyama[1], Yusuke Seto[1],

and Sunao Hasegawa[2]

1: Department of Planetology, Kobe University, Japan

2: Institute of Space and Astronautical Science, Japan Aerospace Exploration Agency, Japan

Contact address:

1-1 Rokkodai-cho, Nada-ku, 657-8501

TEL : +81-78-803-5740

FAX: +81-78-803-5791

e-mail: amnakamu@kobe-u.ac.jp

Abstract

Highly porous small bodies are thought to have been ubiquitous in the early solar system. Therefore, it is essential to understand the collision process of highly porous objects when considering the collisional evolution of primitive small bodies in the solar system. To date, impact disruption experiments have been conducted using high-porosity targets made of ice, pumice, gypsum, and glass, and numerical simulations of impact fracture of porous bodies have also been conducted. However, a variety of internal structures of high-porosity bodies are possible. Therefore, laboratory experiments and numerical simulations in the wide parameter space are necessary.

In this study, high-porosity targets of sintered hollow glass beads and targets made by mixing perlite with hollow beads were used in a collision disruption experiment to investigate the effects of the mixture on collisional destruction of high-porosity bodies. Among the targets prepared under the same sintering conditions, it was found that the targets with more impurities tend to have lower compressive strength and lower resistance against impact disruption. Further, destruction of the mixture targets required more impact energy density than would have been expected from compressive strength. It is likely that the perlite grains in the target matrix inhibit crack growth through the glass framework. The mass fraction of the largest fragment collapsed to a single function of a scaling parameter of energy density in the strength regime ($\Pi_S$) when assuming ratios of tensile strength to compressive strength based on a relationship obtained for ice-silicate mixtures. However, the dependence on $\Pi_S$ is much larger than that shown for porous targets with different internal microstructures from the targets in this study. The depth of the deep cavity specific to the high-porosity target was well represented by a dimensionless parameter using the compressive strength of both the pure glass and mixture targets. The empirical relationship of cavity depth was shown to hold for various targets used in previous studies irrespective of the internal microstructure of the targets.

1. Introduction

Small bodies in the solar system have porous structures. Asteroids with diameters less than tens of kilometers are known to have considerable porosity. This is especially the case for C-class asteroids, which typically have bulk density lower than 2 g cm$^{-3}$ and porosity of 40% or more (Consolmagno et al., 2008). The bulk density and porosity of the Tagish Lake meteorite were reportedly 1.64 g cm$^{-3}$ and ~40%, respectively (Hildebrand, et al., 2006). Small Saturnian satellites have densities from 0.34–0.69 g cm$^{-3}$, which is indicative of a porous internal structure (Porco et al., 2007). The bulk density of comet 67P/Churyumov-Gerasimenko (67P/C-G) is 0.47 g cm$^{-3}$, which corresponds to a porosity of 70–80% if the solid density of the ice-dust mixture is assumed to be 1.5–2.0 g cm$^{-3}$ (Sierks et al., 2015). Comets 19P/Borrelly, 81P/Wild2, and 9P/Tempel1 have porosities in a similar range (Consolmagno et al., 2008).

The uniaxial compressive strength of the near-surface layer of comet 67P/C-G was estimated at > 2 MPa, which suggests that the near-surface ice–dust layer was sintered to have a structure with porosity of 30–65 % (Spohn et al., 2015). Sintering is a microscopic process of mass transfer that physically connects adjacent small particles by forming necks between the particles below the melting temperature. The sintering process depends on temperature. For example, the growth of necks between water ice particles of radius

0.1 µm takes 0.15 years at 100 K (Sirono, 1999). Sintering of water ice particles can progress within the age of the solar system at the radiation equilibrium temperature in the orbit of Jupiter (Gundlach et al., 2018). The temperature condition for sintering of silicate dust particles might also be satisfied for small objects orbiting very close to the Sun: such temperature is ~1000°C for the lunar simulant basalt (Allen et al., 1992) and 1.5 µm amorphous $SiO_2$ particles (Poppe 2003), which is only reached by the bodies within 0.09 AU from the Sun, however, the number of bodies with the perihelion distance smaller than 0.1 AU is very limited in the current era (Granvik et al., 2016). As sintering proceeds, necks grow larger, forming stronger connections between particles (Poppe 2003; Machii and Nakamura, 2011), and the overall structure attains lower bulk porosity.

Impact experiments have been conducted to understand the impact cratering and disruption processes of porous small bodies of various densities and strengths using porous targets consisting of particles physically connected to each other. Sintered glass bead targets with porosity of 5–60 % were impacted by a projectile at impact velocity about 5 km s$^{-1}$ (Love et al., 1993). It was shown that specific energy required to destroy targets greatly depends on porosity. Collisional disruption experiments of sintered glass bead targets with various compressive strength but fixed porosity ~40% at impact velocity between 32 m s$^{-1}$ and 2.2 km s$^{-1}$ showed that the specific energy for disruption increases with the compressive strength of targets (Setoh et al., 2010). Collisional disruption experiments of porous ice and snowball targets with porosity of 30 – 45% and 39 – 54% at impact velocity between 73 and 308 m s$^{-1}$ and 90 and 155 m s$^{-1}$, respectively, showed that the specific energy required for disruption depends on the internal structure and is higher for porous targets than solid ice targets (Ryan et al., 1999; Giblin et al., 2004). Collisional experiments of sintered ice targets including even higher porosity up to 70% at the impact velocity from 2.4 to 489 m s$^{-1}$ confirmed the tendency; more porous target requires more energy density to be catastrophically disrupted (Shimaki and Arakawa, 2012a). On the other hand, it was shown that disruption threshold decreased with increase of porosity in the case of sintered ice-silicate mixture targets with the mass ratio of ice to silicate, 0.5 and with porosity of 0–39% at the impact velocities of 150 to 670 m s$^{-1}$ (Arakawa et al., 2002; Arakawa and Tomizuka, 2004). The reason of the opposite tendency to pure ice target was shown to be due to significant decrease of strength with increasing porosity of the mixture targets.

The number of known low-density, highly porous, small bodies remains limited; however, it is expected that small bodies in the early solar system had high porosity. A theoretical study reported that icy dust grains in the environment of protoplanetary disks accumulate to form planetesimals with radius of 10 km and density of 0.1 g cm$^{-3}$ (Kataoka et al., 2013). A laboratory measurement of the pressure–density relationship of fine particles was extrapolated to estimate the internal porosity structure of granular small bodies. A spherical body with radius of 10 km consisting of particles having the same compression property as 1.7 μm silica beads was shown to have bulk porosity of 82% (Omura and Nakamura, 2018). Numerical simulations of collisional processes have been tested by reproducing laboratory results of non-porous and porous targets for the purpose of effective modeling of collisional processes of small bodies (e.g., Benz and Asphaug, 1994; Jutzi et al., 2009; de Niem, et al., 2018). However, porous small bodies can have a variety of internal structures (Nakamura et al., 2009), so laboratory experiments and numerical simulations covering wide parameter space are useful for understanding the collision process and studying the collisional evolution of small bodies.

To examine the outcome of collisional disruption of targets with high porosity, targets formed of hollow-glass beads with bulk porosity of 87% and 94% have been impacted at velocities between 1.8 and 7 km s$^{-1}$ (Okamoto et al., 2015). The specific energy required for disruption was as large as several kJ/kg, which is much larger than that required for basalt targets. Hollow glass beads are useful for making high porosity structures similar to those of perlite and pumice (Nakamura et al., 2009). The void space of perlite and pumice is formed by evaporation and degassing of volatiles. A similar structure is seen as a feature of scoriaceous cosmic spherules that have lost their volatile components due to heating upon entry into the Earth's atmosphere (Rudraswami et al., 2018). On the other hand, dust grains of sub-micron ~ micron size in protoplanetary disks are considered to form coherent aggregates of very porous structure consisting of filamentous skeleton surrounding void spaces (e.g., Poppe, 2003; Wada et al., 2009). The filamentous framework of the dust aggregate may be simulated by a very thin mesh wall surrounding the void. The hollow glass bead used in a previous study (Okamoto et al., 2015) has thin shell with a thickness of 0.95 μm. Although the shell of the hollow bead is not a mesh structure, a sintered hollow glass bead target may mimic the mechanical and impact response of primitive highly porous bodies formed by the accumulation of porous dust aggregates, especially thermally evolved icy bodies with enhanced bonding between icy dust grains of ~1 μm size. In addition, it has been shown that the structure of silicate dust found in comets varies from solid to very fluffy, including the build-up of sub-structures (Güttler et al., 2019). Accordingly, we conducted impact experiments using the high-porosity targets of hollow glass beads similar to those of the previous study (Okamoto et al., 2015) and targets of porous silicate mixtures to investigate the effects of high porosity and mixture. In this study, experiments were performed only at high impact velocities (> 2 km/s) and compared with previous studies of sintered hollow glass bead targets conducted at similar velocites, however, the low-velocity parameter space needs to be further explored.

2. Experiment

Table 1 summarizes the heating conditions and physical properties of the four different types of targets used in this study. Two of the targets were pure glass bead targets and two were mixtures of glass beads and perlite grains.

The pure glass bead targets with bulk porosities of 86% (HGB87) and 94% (HGB94) and corresponding bulk densities of 0.36 and 0.15 g cm$^{-3}$, respectively, were prepared in a manner similar to that used in previous experiments (e.g., Okamoto et al., 2013). The beads were hollow soda–lime–borosilicate glass microspheres (3M Co.) with an average diameter and shell thickness of 55 μm and 0.95 μm, respectively. The isostatic crush strength of the bead is 5.2 MPa (3M catalogue). This value of strength corresponds to the static uniaxial compressive strength of less-porous (~10% porosity) pure ice (Arakawa and Tomizuka, 2004; Hiraoka et al., 2008). The beads were heated in molds from room temperature to a peak temperature, 800 and 650 °C, over 30 min, respectively. The mold was cup-shaped. We covered the top of the mold with a lid to ensure relatively uniform heating. The peak temperatures were retained for 6 h. Then, the heater was switched off and the targets were cooled to room temperature over 9 h. The HGB87 target had a roughly cylindrical shape with diameter of 58 mm and height of 56 mm. The HGB94 target had roughly the shape of a truncated cone with top and bottom surfaces with diameters of 78 and 65 mm, respectively. The height of this target was 78 mm. The bulk porosity $\phi$ of the target is defined as follows:

$$\phi = 1 - \frac{\rho}{\rho_0}, \tag{1}$$

where $\rho$ is the bulk density of the target and $\rho_0$ is the true density of the constituent material, i.e., 2.5 g cm$^{-3}$ for the glass.

The mixture targets had a mixing ratio of hollow glass beads and perlite grains (<0.6 mm, typically) of 2:1 (mix2:1) and 1:1 (mix1:1) by weight, respectively. Figures 1a and 1b present scanning electron microscopy (SEM) images of the glass beads and perlite grains. The glass beads and perlite grains were placed in a box, shaken for 1 min, poured into molds, and then placed into an oven to be heated to 800 °C. The melting point of perlite is higher than 1093 °C (International Chemical Safety Cards; ICSCs), whereas the softening point of the hollow glass bead material is 600 °C (3M catalogue). The mix2:1 and mix1:1 targets had roughly cylindrical shapes with diameters and heights of 66 mm and 63 mm, and 68 mm and 64 mm, respectively. The bulk porosity of the mixture targets was defined as follows:

$$\phi = 1 - \frac{\rho}{\rho_{01} f_1 + \rho_{02} f_2}, \tag{2}$$

where $\rho_{01}$ and $\rho_{02}$ are the grain densities of the constituent particles and $f_1$ and $f_2$ are the mass fraction of each component, respectively. The perlite grains were themselves porous. We used the true density value of the perlite material (ICSCs), which is 2.2 g cm$^{-3}$. The three targets of HGB87, mix2:1, and mix1:1 were formed with the same heating conditions but with different mixing fraction of perlite grains. The porosity of the target increased with mixing fraction. Figures 1c and 1d present the internal structure of a target and its appearance, respectively. Although the structure is quite inhomogeneous microscopically, the walls of the hollow glass spheres are connected, forming a larger macroscopically continuous structure. The observed thicknesses of necks between glass beads were similar for HGB87, mix2:1, and mix1:1, which is consistent with the fact that they were formed by heating to the same peak temperature. The necks of HGB94 were less thick than the other targets. The microscopic structure of empty void spaces surrounded by thin walls is to some extent similar to the structure of pumice and perlite (Nakamura et al., 2009), although in the case of pumice and perlite the cells are mostly open and the scale of void spaces is larger than in the hollow glass-bead material used in this study.

In order to examine vertical variation of the compressive strength of the target, we sliced the target in the horizontal direction to obtain approximately 2 cm thick discs, from which we cut out three cylinders from the central part of the discs with diameter of 1 cm and length of 2 cm and measured the static uniaxial compressive strength. At some level of compression force, the sample started to locally break and collapse. We defined this force level as the threshold and calculated the threshold force per unit area as the compressive strength. The longitudinal and shear wave velocities were determined by measuring the time required to propagate longitudinal and shear waves through samples of three different thicknesses using piezoelectric sensors.

Impact disruption experiments were conducted at impact velocities of 2.3 to 7.0 km s$^{-1}$ using a two-stage light-gas gun at the Institute of Space and Astronautical Science (ISAS). In high-velocity impact experiments with porous targets, projectile density has been a challenge: lower-density materials generally have smaller strength and we cannot accelerate a projectile with a density comparable to the target density without destroying the projectile. In this study, we successfully launched a wood projectile with a density lower than that of the nylon projectile typically used in previous studies. Nylon spheres with diameter 3.2 mm and wood (Mempisang) columns with diameter of 3 mm and length of ~2.5 mm were accelerated using a split-type nylon sabot (Kawai et al., 2010). Figure 1e presents an image of the wood projectile. The fiber of the wood was parallel to the axis of the column. The compressive strength of the wood column was about 51 MPa, sufficiently strong to be accelerated to more than 5 km s$^{-1}$. The strength of the column was also measured by pressing from the side of the column, which corresponds to the Brazilian disc test (Diyuan and Louis, 2013). The tensile strength of the column was thus determined to be 2.4 MPa. Table 2 summarizes the physical properties of the projectiles.

The targets were either hung by thread or placed on a stand made of plastic or paper. The bottom and backside wall of the experimental chamber were covered with plastic cushioning. Nominally the target was set with its symmetry axis aligned with the projectile trajectory so that it would be impacted on the top surface. However, in the case of the target being hung by thread, the attitude of the target was random and the projectile could strike obliquely onto the top, side, or even bottom surface of the target. To check the point of impact, we used two high-speed video cameras to acquire imagery from directions orthogonal to the projectile trajectory. Shimazu HPV-X and HPV-1 cameras were used in earlier shots, and the Shimazu HPV-X and a Kirana-05M were used in later shots. The cameras were nominally operated at $(2–5) \times 10^5$ fps with exposure durations 0.2–1 and 0.5 μs, respectively. The motion of the largest fragments was monitored using another high-speed video camera, the Photron SA1.1, operated at 5400 fps with an exposure duration of 50-185 μs. Additionally, because the projectile was accelerated using a sabot, the trajectory could not be adjusted accurately toward the center of mass of the target. The impact parameter $b$ was defined as:

$$b = \frac{h}{R}, \tag{3}$$

where $h$ is the distance between the target's center of mass and the projectile trajectory and $R$ is the radius of the sphere having the same volume as the target (equivalent sphere radius). We defined an apparent impact angle $\theta$, which is the angle between the projectile trajectory and the impacted surface of the target looking from the vertical direction (i.e., looking down from the top), for example, $\theta=90°$ indicates an apparent normal impact on the surface. Because the projectile could strike on the side surface of the target higher or lower than the target's center of mass, the apparent impact angle is not an exact one. Table 3 summarizes the impact conditions.

3. Result

3.1 Static strength and wave velocities

Figure 2 presents the measurement results of compressive strength, $Y_c$. The horizontal axis of Fig. 2a indicates the depth of the cylindrical core specimen normalized by the height of the target. The strength of the parts near the top and the bottom surfaces was generally lower than that of the central part. The variation within a target was most prominent in the most-porous target HGB94, probably because it had the lowest thermal conductivity: the temperature in the central part of the target would have declined more slowly than the temperature near the surface of the target. The variation was within a factor of 2, and we hereafter refer to the average value of strength. Figs. 2b compares the compressive strength and bulk density of the targets in this study with those in previous studies. The compressive strength of HGB87 (1.7 ± 0.4 MPa) agreed to within 1 $\sigma$ with a target prepared using the same peak temperature and duration in a previous study, which had a compressive strength of 1.4 ± 0.4 MPa (fluffy87 in Okamoto et al., 2013). The compressive strength of HGB94 (0.09 ± 0.03 MPa) was lower than one with the same porosity, which had a compressive strength of 0.47 ± 0.13 MPa (fluffy94 in Okamoto et al., 2013), because of the lower peak temperature used during the heating process in this study. The compressive strength decreased with the increase of mixing fraction of perlite grains, i.e., HGB87 > mix2:1 > mix1:1. The wave velocity had a tendency similar to that of the compressive strength: the higher the perlite fraction, the slower the wave velocity. In other words, the targets formed using the same peak temperature (HGB87, mix2:1, and mix1:1) had higher compressive strengths and wave velocities with increases in the fraction of glass and bulk density. Table 1 lists the measurement results.

## 3.2 $Q_S^*$

Figure 3 presents an example of a set of fragments collected after impact. In most of the shots, a bulb-shaped cavity formed below the impact point and the target separated into a few to tens of larger pieces. The largest fragments moved with a velocity lower than several m s$^{-1}$ in a direction almost parallel to the projectile trajectory. The center-of-mass velocity of the system was between 0.9 and 2.6 m s$^{-1}$ and the velocities of the largest fragments were within the same order of magnitude. Most of them were not broken when struck and landed on the cushioning.

Figure 4 presents the results of the largest fragment mass fraction of the target, $M_1/M$, where $M_1$ and $M$ are the mass of the largest fragment and the initial mass of the target, respectively, versus the specific energy of impact $Q$ which is defined as

$$Q = \frac{mV^2}{2(M+m)}, \qquad (4)$$

where $V$ and $m$ are the impact velocity and mass of projectile, respectively. The results of HGB87 agree with the previous results of fluffy87 (Okamoto et al., 2015), although our data here include data for various impact angle $\theta$ and impact parameter $b$ values, whereas the previous data were obtained only for $\theta = 90°$. No information about $b$ for the previous data is available. The projectiles used in the previous study were nylon and titanium spheres: therefore, the data collectively show no clear difference between projectiles of different density ranging from the value of 0.74 g cm$^{-3}$ of wood to the value of 4.5 g cm$^{-3}$ of titanium. Such insensitivity to projectile material was also observed in a previous study of rock disruption (Katsura et al., 2014).

As expected from the lower static strength of HGB94 than fluffy94, the HGB94 target was easier to destroy than the fluffy94 target. The data of the mix2:1 and mix1:1 targets were more scattered than those of the HGB87 and HGB94 targets, although the all target types had similar degree of scatter in the static strength (24-32 %) as shown in Table 1, i.e., variations in the bulk properties of the targets by the manufacturing process were similar. The data of mixture targets are plotted between those of HGB87 and HGB94, which is in agreement with the fact that the mix2:1 and mix1:1 targets had static strength between those of the HGB87 and HGB94 targets. The scattering of the data of the mixture targets is probably due to the inhomogeneity in internal structure of these targets. Crack growth in homogeneous targets is reproducible, whereas crack growth in mixture targets has lower reproducibility. Crack growth in a hollow glass bead structure stops when the crack crosses the boundary between the hollow glass bead and a perlite grain.

Least-square fits of the following equations were applied to the data:

$$\frac{M_1}{M} = a_1 Q^{-b_1}, \tag{5}$$

$$Q = a_2 \left(\frac{M_1}{M}\right)^{-b_2} \tag{6}$$

and Table 4 lists the fitted parameters. We derived the shattering specific energy $Q_S^*$ using the parameter of Eq. 6 required to make the largest fragment have half the mass of the initial target, i.e., $\frac{M_1}{M} = 0.5$. The ambiguity of the estimated $Q_S^*$ was calculated according to error propagation. We confirmed that values of $Q_S^*$ derived from Eq. 5 agree with those from Eq. 6 within the ambiguity shown in Table 4. The fittings were applied to the data irrespective of $\theta$ and $b$. Figure 4 presents the fitting curves. To summarize, the sintering condition of HGB87, mix2:1, and mix1:1 was the same. The resultant compressive strength and specific energy $Q_S^*$ were HGB87 > mix2:1 > mix1:1 and HGB87 > mix2:1 ≥ mix1:1, respectively, as shown in Table 1 and Table 4, i.e., more fraction of impurity, less static strength and impact resistance.

3.3 Cavity

The results revealed that the cavity below the impact point in a highly porous target becomes bulb-shaped when the projectile is broken (Okamoto et al., 2013; 2015). Figure 5 presents an example of a cavity. We measured the depth of cavity $d_b$ and the depth at the maximum diameter of cavity $d_c$, which would correspond to the depth of the center of the spherical source of shock-wave divergence (the center of the isobaric core) (Mizutani et al., 1990), using images of the fragments recovered after shots. Because the symmetry axis of the bulb is not necessarily in the fracture surface, the values shown in Table 5 are reference values. The range of the ratios of these measurements, $\frac{d_b}{d_c}$, was 1.7 ± 0.3. The depth of cavity in this study was as deep as roughly half the target, i.e., $\frac{d_b}{L_t} \sim 0.5$, where $L_t$ denotes the height of the target.

4. Discussion

4.1 Effects of impact parameter and oblique incidence

As mentioned in section 3.2, the impact parameter and impact angle had minor effects on the largest-fragment mass fraction. This tendency is different from what was found in previous impact disruption experiments conducted using spherical rock targets (Fujiwara and Tsukamoto 1980; Nakamura 1993). Note that in the case of a spherical target, the impact parameter has a one-to-one correspondence to the impact angle. Figure 6 presents $Q/Q_{effct}$ versus impact parameter and impact angle, where $Q_{effect}$ is the calculated value required to obtain the measured largest-fragment mass $M_1$ of each shot using Eq. (5). Figure 6 also presents previous results for basalt spheres. In the case of basalt spheres, an effect of impact geometry appears at $b > \sim 0.5$ and $\theta < 60°$: however, the results of this study show no clear effect. The degree of destruction was probably insensitive to the impact geometry because of the target shape and the high porosity of the target. The mass of fragments excavated by an impact is known to be dependent on the curvature of the target surface (Fujiwara et al., 1993; Fujiwara et al., 2014; Suzuki et al., 2018) and the distance between the point that corresponding to the center of the isobaric core of pressure propagation and the target free surface (Suzuki et al., 2018). In the present case, the roughly cylindrical shape of the targets and their highly porous internal structure collectively made the distance from the center of the isobaric core to the free surface insensitive to collision geometry, and thus the outcome was insensitive to collision geometry. Additional experiments with spherical targets can reveal the geometric effect of the collision more clearly.

4.2 Cavity depth

The cavity depth depends on the target type: deeper depths for more fragile, lower-density targets. A previous study proposed an empirical relationship between depth at maximum cavity width, $d_c$ in this study, and $L_0 \equiv \frac{1}{2\alpha}$ as:

$$\frac{d_c}{d} = 10^{0.60}(\frac{L_0}{d})^{0.46}, \qquad (7)$$

Where $d$ is the projectile diameter and $\frac{1}{2\alpha}$ is the characteristic distance through which the projectile is decelerated in the porous target:

$$m\frac{dV}{dt} = -\frac{1}{2}C_d \rho S V^2, \qquad (8)$$

$$\alpha \equiv \frac{C_d \rho S}{2m}, \tag{9}$$

where $C_d$ and $S$ are the drag coefficient and projectile cross-sectional area, respectively (Okamoto et al., 2015). When the projectile deforms or breaks, the cross-sectional area and the mass of the projectile change from the initial values. Using X-ray transmission images, the following empirical relationship was derived:

$$\frac{d}{L_0} = \frac{3}{2} C_d \left(\frac{\rho}{\delta}\right) = \frac{3}{2} \times 10^{-0.039} \left(\frac{\rho V^2}{Y_{pt}}\right)^{0.22} \left(\frac{\rho}{\delta}\right), \tag{10}$$

where $\delta$ is the projectile density and $Y_{pt}$ is the tensile strength of projectile. Figure 7 presents $\frac{d_c}{d}$ versus $\frac{L_0}{d}$ of this study. We adopted the 'compressive' strength of the wood projectile instead of the 'tensile' strength taking the direction of acceleration into account. The results are not as consistent with Eq. (7) as were seen for fluffy 87 and fluffy 94 in the previous study. It reveals a larger dependence on the property of the target, i.e., density or compressive strength.

To take the target strength $Y$ into account, we adopted the non-dimensional forms according to pi-group scaling (Holsapple, 1993):

$$\pi_{db}/\pi_4^{\beta_{db}} \propto \pi_3^{\alpha_{db}}, \quad \pi_{dc}/\pi_4^{\beta_{dc}} \propto \pi_3^{\alpha_{dc}} \tag{11}$$

where $\alpha_{db}$, $\alpha_{dc}$, $\beta_{db}$, and $\beta_{dc}$ are fitting parameters. The non-dimensional parameters are defined as,

$$\pi_{db} = \left(\frac{\rho}{m}\right)^{1/3} d_b, \quad \pi_{dc} = \left(\frac{\rho}{m}\right)^{1/3} d_c, \quad \pi_3 = \frac{Y}{\delta V^2}, \quad \pi_4 = \frac{\rho}{\delta}. \tag{12}$$

We assumed $\beta_{db} = \beta_{dc} = 0.01$ according to the value of $\beta_{db}$ obtained in a previous study of a sedimentary rock with porosity of ~17% (Suzuki et al., 2012). We adopted the compressive strength $Y_c$ as $Y$ in this study, whereas the previous study adopted the tensile strength as $Y$ (Suzuki et al., 2012). Figure 8a presents the results including those of hollow glass bead targets from a previous study (Okamoto and Nakamura, 2017). Although the previous ones are data of cratering shots, i.e., the targets were not broken as a whole, and the targets in this study were destroyed, the depth data appear similar. Contrary to the result of $\frac{M_1}{M}$, the scattering of $\pi_{db}$ of the mixture targets is not apparently larger than those of pure targets. This probably occurred because the cavity depth depends on the compressibility of the target, which is characterized by the compressive strength $Y_c$. We obtained empirical relationships by least squares fittings to the results of this study:

$$\frac{\pi_{db}}{\pi_4^{0.01}} = (1.057 \pm 0.030)\pi_3^{-0.181 \pm 0.035}, \qquad (13)$$

and

$$\frac{\pi_{dc}}{\pi_4^{0.01}} = (0.158 \pm 0.006)\pi_3^{-0.303 \pm 0.045}. \qquad (14)$$

The density ratios of projectile to target in this study were between 2 and 7. Ordinary chondrites (3.2 ~ 3.4 g cm$^{-3}$) (Consolmagno et al., 2008) and comets (for example, 0.47 g cm$^{-3}$ of 67P/C-G) reach a density ratio of about 7, but collisions between objects with a density ratio of about unity also occur in interplanetary space.

To anticipate the outcome of a collision of equally dense objects, we examined whether the empirical relationship shown by Eq. (13) could be extrapolated to the case of collision between projectile and target with smaller density ratio. Figure 8b compares the results of this study with the results of previous studies with projectile to target density ratio less than 2. Pumice and gypsum targets (0.59 and 1.1 g cm$^{-3}$) were impacted by a nylon projectile (1.1 g cm$^{-3}$) with velocities between 3.6 and 7.2 km s$^{-1}$ and a velocity fixed at 3.5 km s$^{-1}$, respectively (Okamoto and Nakamura, 2017; Kadono et al., 2018). Data of a cratering experiment of snow target (36% porosity) conducted with velocities between 27 and 145 m s$^{-1}$ are also shown (Arakawa and Yasui, 2011). We assumed that the dynamically determined strength of snow is comparable to that of the statically measured compressive strength of other materials. All of the previous data of the variety of materials clustered near the empirical relationship of Eq. (13). Shallow craters in the case of snow projectiles rather than ice projectiles may be explained by the lower strength of the snow projectiles. Figure 8b presents the result of crater depth on a basalt target for which the projectile to target density ratio was 0.9 (Table A-1) and an empirical relationship obtained for sedimentary rock (Suzuki et al., 2012), too. The result for basalt falls roughly on the line of Eq. (13). The discrepancy in the empirical relationship for the sedimentary rock target compared with the other results is probably due to the use of the target's tensile strength instead of the target's compressive strength. In summary, the empirical relationship (Eq. 13) holds within factor of 2 not only for the present results but also for the experimental results with projectile to target density ratio between 0.9–2, irrespective of the internal microstructure of the targets: the sintered hollow glass bead targets of this study and pumice target have the porous structure consisting of void spaces surrounded by walls, whereas gypsum and snow targets are coherent aggregates (Nakamura et al., 2009).

4.3 Effect of mixture on degree of destruction

As described in section 3.2, Fig. 4 shows that $Q_s^*$ depends on target strength. Yasui and Arakawa (2011) found that the results of impact disruption experiments of gypsum and gypsum–glass bead mixed targets, which simulated the parent bodies of ordinary chondrites, were well organized using a non-dimensional parameter, that is, the non-dimensional impact stress, $P_{It}$, which is defined as (Mizutani et al., 1990):

$$P_{It} = \frac{V_{projectile}}{V_{target}} \frac{P_0}{Y_t}, \qquad (15)$$

where $P_0$, $V_{projectile}$, $V_{target}$, and $Y_t$ are initial shock pressure, volume of projectile, volume of target, and tensile strength of target material, respectively. In this study, the tensile strength of target was not measured, so we tried a modified version of impact stress,

$P_{Ic}$ (Takagi, et al., 1984; Mizutani et al., 1990) which is defined as:

$$P_{Ic} = \frac{V_{projectile}}{V_{target}} \frac{P_0}{Y_c}. \tag{16}$$

We calculated $P_0$ using Hugoniot parameters and the planar approximation (Melosh, 1989). The shock wave velocity $U_i$ and $P_0$, where $i=p$ and $t$ are projectile and target, respectively, are expressed as:

$$P_0 = \rho U_t u_{pt} \; or \; \delta U_p u_{pp}, \tag{17}$$

$$U_i = C_i + s_i u_{pi}, \tag{18}$$

where $C_i$, $s_i$, $u_{pi}$ are bulk sound velocity, material constant, and particle velocity, respectively. Table 6 summarizes the Hugoniot parameters $C$ and $s$ we used. Figure 9a shows that the results of the pure glass targets (HGB87 and 94) collapse into a narrower range than in the case in which specific energy $Q$ was used (Fig. 4). In contrast, the results of the mixture targets show higher resistance to impact destruction than that expected based on the compressive strength of the target material.

The definitions of $P_{It}$ and $P_{Ic}$ as expressed in Eqs. (15) and (16), respectively, are approximations when the depth of the center of the isobaric core is negligible compared to the target length, i.e., $\frac{d_c}{L_t} \ll 1$. However, in the present study, $\frac{d_c}{L_t} \sim \frac{1}{1.7} \times 0.5 \sim 0.3$ and may not be negligible, so we took the effect of the depth of the center of the isobaric core into account. Figure 9b presents the largest-fragment mass fraction and the modified version of impact stress $P_{Ic}'$ for the shots in which the projectile impacted the top or bottom surface of the target with impact angle $\theta > 80°$;

$$P_{Ic}' = \frac{V_{projectile}}{(L_t - d_c)^3} \frac{P_0}{Y_c}. \tag{19}$$

We used values of $d_c$ calculated using the empirical relationship expressed in Eq. (14). Because of the low compressive strength of the target, the depth of the center of the isobaric core of the mixture target is large: thus, the difference between the pure glass target and the mixture target does not become small by this modification.

The apparent high resistance of the mixture targets shown in Figs. 9a and 9b may be interpreted as the fracture growth in the network of glass beads in the mixture target being blocked by perlite grains or by pores introduced by the mixing with perlite grains, making it difficult for fracture to propagate. Not only the structural discontinuities introduced by perlite grains, but also the relative weakness of perlite grains over hollow glass beads could have acted as an obstacle for the propagation of fracture. In other words, more energy may have dissipated by the pulverization of perlite grains. Hiraoka et al. (2008) found that the ratio of tensile strength to compressive strength of an ice–silicate mixture with porosity of ~10% increased with increase of silicate fraction, $f$ ($100 \times f\%$) as:

$$\frac{Y_t}{Y_c} = \frac{0.68 e^{2.6f}}{6.5 e^{0.4f}} = 0.10 e^{2.2f}. \tag{20}$$

The tendency of the silicate mixture to increase the ratio of tensile strength to compressive strength of the mixture target is consistent with the results shown in Figs. 9a and 9b.

A non-dimensional scaling parameter $\Pi_{s\_original}$, which was introduced based on the coupling parameter concept (Holsapple and Housen, 1986; Housen and Holsapple, 1990), was shown to be effective for compiling laboratory results of targets with various porosities conducted at a wide range of velocities (Okamoto et al., 2015). By using this parameter, the largest-fragment mass fraction data of gypsum and gypsum–glass bead mixed targets were fitted by a single relationship (Yasui and Arakawa, 2011). The expression of $\Pi_{s\_original}$ is (Eqs. 40 of Housen and Holsapple, 1990):

$$\Pi_{s\_original} = Q_s \left(\frac{\rho}{Y}\right)^{\frac{3\mu}{2-\lambda}} R^{-3\mu(\lambda+\tau)/(\tau-2)} V^{3\mu-2} \left(\frac{\rho}{\delta}\right)^{1-3\nu}, \quad (21)$$

where $R$ and $Y$ denote target radius and strength, respectively, $\lambda$ and $\tau$ describe dependence of the strength of target material on size scale and strain rate, respectively, and $\mu$ and $\nu$ are constants used in the coupling parameter. When the strength of the target material does not depend on size scale but only on the strain rate ($\lambda = 0$) and if we assume $\nu = \frac{1}{3}$ then we obtain:

$$\Pi_{s\_original}(\lambda = 0, \nu = \tfrac{1}{3}) = R^{-3\mu\tau/(\tau-2)} Q_s \left(\frac{\rho}{Y}\right)^{\frac{3\mu}{2}} V^{3\mu-2}. \quad (22)$$

The value of $\mu$ is theoretically limited between 1/3 and 2/3. Similarly to the case of the previous study (Okamoto et al., 2015), we assumed a linear relationship between $\mu$ and porosity, $\mu = \frac{2}{3} - \frac{\phi}{3}$. Thus, the value of $\mu$ becomes 0.35–0.38 for glass beads and mixture target materials and 0.42–0.43 for pumice (Table A-2), which are similar to those of porous targets such as perlite–sand mixture with 60% porosity ($\mu = 0.35$), sand–fly ash mixture with 45% porosity ($\mu = 0.4$) (Housen and Holsapple, 2011), and gypsum with 65–69% porosity ($\mu = 0.398$) (Nakamura et al., 2015). Figure 10a shows $\frac{M_1}{M}$ versus $Q_s(\frac{\rho}{Y_c})^{\frac{3\mu}{2}}V^{3\mu-2}$ for glass bead and mixture targets as well as pumice data. We substituted $Y_c$ for $Y$ of Eq. (22). The data of the glass beads (HGB87 and HGB94) and the pumice are fitted by a single relationship:

$$\frac{M_1}{M} = (2.5 \pm 0.7) \times 10^{-4} \{Q_s(\frac{\rho}{Y_c})^{\frac{3\mu}{2}}V^{3\mu-2}\}^{-1.52\pm0.07}. \qquad (23)$$

The agreement between the hollow glass bead and pumice results may be due to the similar internal structure of the two substances. The discrepancy of the data of the mixture targets from Eq. (23) is probably due to the effect of the perlite grains on crack growth in the mixture targets. We used Eq. (20) to anticipate the tensile strength of the targets used in this study, although the ratio of tensile strength to compressive strength was obtained for different compositions (ice and serpentinite powders in the previous study) and for different porosity (~10%). For pumice, we used the measured value (1 MPa, Jutzi et al., 2009). In Fig. 10b, all the data are collapsed into a single relationship:

$$\frac{M_1}{M} = (1.16 \pm 0.04) \times 10^{-2} \Pi_s^{-1.68\pm0.07}. \qquad (24)$$

where,

$$\Pi_s = Q_s \left(\frac{\rho}{Y_t}\right)^{\frac{3\mu}{2}} V^{3\mu-2}. \tag{25}$$

The fitted line is shown in the figure. The power index of $\Pi_s$ in Eq. (24) is $-1.7$ and shows much larger dependence than that found for the gypsum and gypsum–glass bead mixed targets ($-0.96$) (Yasui and Arakawa, 2011). The cavity depth is expressed by the power law relationship of the dimensionless forms (Eq. 13) irrespective of the internal microstructure of the target, while the largest fragment mass fraction has different dependence on $\Pi_{s,t}$ depending on the internal microstructure of the target.

5. Summary

There were likely a variety of porous small bodies in the solar system, especially in the early phase of the solar system. Laboratory and numerical studies on the collisional processes of porous bodies of various internal structure increase our understanding of these collisional process. In this study, we conducted impact disruption experiments of highly porous targets made of pure glass and glass–silicate mixture. The targets have a porosity in the range of 86 ~ 94%, and may be regarded as a mimic of primitive highly porous bodies consisting of dust aggregates with enhanced bonding due to thermal processing. The impact velocities ranged from 2.3–7.1 km s$^{-1}$ in this study, however, to understand the collisional evolution of small bodies since the early solar system, it is also necessary to study the low-velocity parameter space. To reduce the density ratio of projectile to target, we used nylon and wood projectiles.

The results revealed that the depth of the deep cavity specific to the high-porosity target was well represented by the dimensionless parameter of PI-group scaling for cratering in the strength regime using the compressive strength of both the pure glass targets and the mixture targets. It was shown that the same relationship holds widely for other targets with different internal microstructures. The largest fragment mass fraction was insensitive to collision geometry, i.e., impact parameter and impact angle, probably due to the deep center of the isobaric core. The targets with more impurities tend to have lower compressive strength and lower resistance against impact disruption. Further, the mixture targets required more impact energy density than would have been expected from the static compressive strength. This was probably because the impurity inhibited the growth of cracks in the framework structure made of glass. The largest-fragment mass fraction of the pure glass targets and the mixture targets, as well as the results of pumice targets, collapsed to a single function of a non-dimensional scaling parameter of energy density in the strength regime ($\Pi_S$) (e.g., Housen and Holsapple, 1990) by assuming ratios of tensile strength to compressive strength based on a relationship obtained for an ice–silicate mixture (Hiraoka et al., 2008). The largest fragment mass fraction obtained in this study showed a greater dependence on $\Pi_S$ than previously obtained for porous targets with different internal microstructures.

The tensile strength of a mixture depends on various factors, including porosity, fraction of impurity, composition of impurity, temperature, and grain size (e.g., Arakawa and Tomizuka, 2004; Hiraoka et al., 2008; Litwin et al., 2012). Although the results of this study suggested that the degree of collisional destruction of the targets would depend on tensile strength rather than compressive strength, the tensile strength of the targets were not directly measured in this study. Further studies on the tensile strength of high-porosity structures of primitive small bodies in the solar system are needed, including whether the tensile strength of the targets used in this study has the impurity fraction dependence assumed here.

Acknowledgement

We are grateful to the constructive comments of two anonymous reviewers. The series of experiments was supported by the Hypervelocity Impact Facility (the Space Plasma Laboratory) at ISAS, raJAXA, Japan. Y. M. is grateful to M. Hyodo for allowing him to acquire SEM images of the hollow-glass spheres and perlite grains.

Appendix

Table A-1 lists the conditions of the impact experiment with the basalt (Kinosaki, Japan) target. The target was a rectangular parallelepiped with sides ranging from 12.1 cm to18.5 cm. The density of the target was 2.7 g cm$^{-3}$. We did not measure the strength of the target material and we assumed the compressive strength of 220 MPa according to a previous study (Takagi et al., 1984). Crater depth was measured using a laser profiler and is shown in the table (e.g., Suzuki et al., 2012). Table A-2 lists the conditions of the impact experiment with the pumice target. A portion of the result was presented in previous studies (Nakamura et al., 2009; Jutzi et al., 2009): however, a few new experiments were conducted after these studies.

Table 1 Sintering conditions and physical properties of targets.

| Type | Peak temp. (°C) | Height $L_t$ (mm) | Diameter[1] (mm) | Bulk density, $\rho$ (g cm$^{-3}$) | Porosity $100 \times \phi$ (%) | Compressive strength, $Y_c$ (MPa) | Longitudinal wave speed, $V_p$ (km s$^{-1}$) | Shear wave speed, $V_s$ (km s$^{-1}$) |
|---|---|---|---|---|---|---|---|---|
| HGB87 | 800 | 56 | 58 | 0.36±0.01 | 85.8 | 1.70±0.43 | 1.49±0.04 | 1.07±0.05 |
| HGB94 | 650 | 78 | 78 (t) 65 (b) | 0.15±0.00 | 94.1 | 0.088±0.028 | 0.71±0.01 | 0.43±0.03 |
| Mix2:1 | 800 | 63 | 66 | 0.26±0.01 | 89.3 | 0.25±0.06 | 0.70±0.04 | 0.325±0.002 |
| Mix1:1 | 800 | 64 | 68 | 0.23±0.01 | 90.6 | 0.092±0.022 | 0.57±0.02 | 0.33±0.01 |

1) t: top surface, b: bottom surface.

Table 2 Properties of projectiles.

| Type | Shape | Size (mm) | Density $\delta$ (g cm$^{-3}$) | Strength (MPa) |
|---|---|---|---|---|
| Nylon | Sphere | 3.175 (diam.) | 1.1 | 62 – 83[1] |
| Wood | Column | 3 (diam.)×2.5 (length) | 0.74 | 2.4 (tensile) 51 (compressive) |

1) Chronological Scientific Tables (2019).

Table 3 Impact conditions and the largest-fragment mass fraction.

| Shot # | Target | | Projectile | | | Target support [2] | f [3] | θ [4] (°) | b | $M_1/M$ |
|---|---|---|---|---|---|---|---|---|---|---|
| | Type | M (g) | Type [1] | m (g) | V (km s$^{-1}$) | | | | | |
| m11 | HGB87 | 53.994 | w | 0.013 | 2.60 | thread | t | 90 | 0.30 | 0.9978 |
| m10 | HGB87 | 54.820 | w | 0.016 | 2.60 | thread | s | 63 | 0.54 | 0.9931 |
| m13 | HGB87 | 54.406 | w | 0.013 | 3.32 | thread | s | 35 | 0.52 | 0.9902 |
| m17 | HGB87 | 51.938 | w | 0.013 | 4.78 | thread | b | 56 | 0.04 | 0.9351 |
| m15 | HGB87 | 55.682 | w | 0.017 | 5.17 | thread | s | 86 | 0.90 | 0.9387 |
| m14 | HGB87 | 50.860 | w | 0.014 | 5.27 | thread | s | 28 | 1.0 | 0.9303 |
| m2 | HGB87 | 53.219 | n | 0.019 | 5.07 | stand1 | t | (90) | 0.32 | 0.7815 |
| m12 | HGB87 | 52.094 | n | 0.019 | 5.09 | thread | s | 35 | >0.04 | 0.7021 |
| m25 | HGB87 | 55.870 | n | 0.019 | 5.23 | thread | s | 53 | >0.56 | 0.8124 |
| m24 | HGB87 | 55.413 | n | 0.019 | 6.20 | thread | s | 42 | 0.25 | 0.5013 |
| m8 | HGB87 | 50.959 | n | 0.019 | 6.30 | stand2 | t | (90) | 0.25 | 0.3618 |
| m6 | HGB87 | 51.712 | n | 0.019 | 6.31 | stand2 | t | (90) | 0.28 | 0.5405 |
| m5 | HGB87 | 51.958 | n | 0.019 | 7.02 | stand2 | t | (90) | 0.31 | 0.3394 |
| m3 | HGB87 | 53.694 | n | 0.019 | 7.02 | stand1 | t | (90) | 0.27 | 0.3165 |
| m4 | HGB87 | 55.467 | n | 0.019 | 7.09 | stand2 | t | (90) | 0.50 | 0.4021 |
| m26 | HGB94 | 51.194 | w | 0.012 | 2.30 | stand2 | t | (90) | 0.06 | 0.9713 |
| m30 | HGB94 | 47.047 | w | 0.010 | 2.58 | stand2 | t | (90) | - | 0.9764 |
| m33 | HGB94 | 47.574 | w | 0.011 | 3.03 | stand2 | t | (90) | 0.47 | 0.9949 |
| m32 | HGB94 | 45.824 | w | 0.012 | 3.54 | stand2 | t | (90) | 0.71 | 0.6370 |
| m29 | HGB94 | 47.387 | w | 0.011 | 5.46 | stand2 | t | (90) | 0.38 | 0.3219 |
| m27 | HGB94 | 46.223 | n | 0.019 | 5.24 | stand2 | t | (90) | 0.49 | 0.08434 |
| m40 | mix2:1 | 53.082 | n | 0.019 | 3.13 | thread | t | 21 | 0.42 | 0.6990 |
| m48 | mix2:1 | 55.956 | n | 0.019 | 3.50 | thread | s | 64 | 0.32 | 0.9552 |
| m50 | mix2:1 | 56.579 | n | 0.019 | 4.00 | thread | s | 42 | 0.28 | 0.8655 |
| m39 | mix2:1 | 54.643 | n | 0.019 | 4.15 | thread | s | 59 | 0.54 | 0.7430 |
| m46 | mix2:1 | 54.918 | n | 0.019 | 4.20 | thread | t | 87 | 0.34 | 0.4395 |
| m41 | mix2:1 | 56.070 | n | 0.019 | 4.28 | thread | t | 87 | 0.49 | 0.6520 |
| m53 | mix2:1 | 55.527 | n | 0.019 | 5.00 | stand2 | t | (90) | 0.27 | 0.5498 |
| m34 | mix2:1 | 55.400 | n | 0.019 | 5.11 | thread | s | 65 | 0.42 | 0.3450 |
| m42 | mix1:1 | 52.649 | w | 0.011 | 4.79 | thread | s | 75 | 0.41 | 0.9862 |
| m54 | mix1:1 | 49.540 | w | 0.012 | 4.99 | stand2 | t | (90) | 0.49 | 0.8236 |
| m51 | mix1:1 | 49.805 | w | 0.011 | 5.05 | stand2 | t | (90) | 0.52 | 0.8651 |
| m47 | mix1:1 | 54.056 | n | 0.019 | 3.76 | thread | s | 62 | 0.16 | 0.4809 |
| m55 | mix1:1 | 52.688 | n | 0.019 | 3.87 | stand2 | t | (90) | 0.12 | 0.5280 |
| m49 | mix1:1 | 52.180 | n | 0.019 | 4.07 | thread | t | 66 | 0.26 | 0.4065 |
| m52 | mix1:1 | 54.479 | n | 0.019 | 4.30 | stand2 | t | (90) | 0.29 | 0.4594 |
| m45 | mix1:1 | 54.600 | n | 0.019 | 4.39 | thread | b | 86 | 0.27 | 0.2547 |
| m43 | mix1:1 | 55.370 | n | 0.02 | 5.09 | thread | s | 90 | 0.61 | 0.4586 |
| m44[5] | mix1:1 | 52.461 | n | 0.019 | 6.10 | thread | b | - | >0.19 | 0.2784 |

1) w: wood cylinder, n: nylon sphere

2) Method of supporting target. "thread": target was hung by thread. "stand1": target was put on a plastic stand. "stand2": target was put on a paper stand.

3) The face of target hit by the projectile. "t": top. "s": side. "b": bottom.

4) In case of use of a stand for target support, we did not measure the impact angle, however it is expected almost normal impact (impact angle ~ 90°).

5) Images of one of cameras was not available.

Table 4 Fitted parameters of Eqs. (5) and (6), and the disruption threshold $Q_s^*$.

| Target | $a_1$ | $b_1$ | $a_2$ | $b_2$ | $Q_s^*$ |
|---|---|---|---|---|---|
| HGB87 | $2.31(\pm 0.06) \times 10^4$ | $1.22 \pm 0.11$ | $3.85(\pm 0.11) \times 10^3$ | $0.77 \pm 0.07$ | $6.54(\pm 0.33) \times 10^3$ |
| HGB94 | $1.69(\pm 0.11) \times 10^4$ | $1.39 \pm 0.26$ | $1.12(\pm 0.11) \times 10^3$ | $0.67 \pm 0.13$ | $1.78(\pm 0.20) \times 10^3$ |
| Mix2:1 | $2.57(\pm 0.22) \times 10^2$ | $0.75 \pm 0.31$ | $2.19(\pm 0.16) \times 10^3$ | $0.66 \pm 0.27$ | $3.44(\pm 0.67) \times 10^3$ |
| Mix1:1 | $6.6(\pm 0.7) \times 10^2$ | $0.88 \pm 0.41$ | $2.55(\pm 0.25) \times 10^3$ | $0.42 \pm 0.19$ | $3.42(\pm 0.54) \times 10^3$ |

Table 5 Cavity depth.

| Shot # | Target type | Projectile type | $d_c$ (mm) | $d_b$ (mm) | $d_b/L_t$ |
|---|---|---|---|---|---|
| m2 | HGB87 | n | 11.4 | 28.1 | 0.502 |
| m3 | HGB87 | n | 14.2 | 25.3 | 0.452 |
| m4 | HGB87 | n | 14.6 | 24.3 | 0.434 |
| m5 | HGB87 | n | 12.7 | 19.8 | 0.354 |
| m6 | HGB87 | n | 9.5 | 21.3 | 0.380 |
| m8 | HGB87 | n | 10.5 | 22.1 | 0.395 |
| m32 | HGB94 | w | 27.0 | 43.6 | 0.559 |
| m39 | mix2:1 | n | 21.8 | 37.0 | 0.587 |
| m40 | mix2:1 | n | 19.2 | 31.6 | 0.502 |
| m41 | mix2:1 | n | 22.3 | 35.2 | 0.559 |
| m46 | mix2:1 | n | 20.6 | 39.2 | 0.622 |
| m53 | mix2:1 | n | 24.8 | 38.2 | 0.606 |
| m43 | mix1:1 | n | 30.9 | 44.2 | 0.691 |
| m45 | mix1:1 | n | 18.3 | 33.9 | 0.530 |
| m51 | mix1:1 | w | 24.4 | 36.0 | 0.563 |
| m52 | mix1:1 | n | 26.3 | 41.1 | 0.642 |
| m54 | mix1:1 | w | 23.0 | 35.0 | 0.547 |
| m55 | mix1:1 | n | 31.6 | 42.0 | 0.656 |

Table 6 Parameters used in this study.

| Material | Density (g cm$^{-3}$) | $C$ (km s$^{-1}$) | $s$ |
|---|---|---|---|
| HGB87 | 0.36[1] | 0.83[2] | 1.4[3] |
| HGB94 | 0.15[1] | 0.51[2] | 1.4[3] |
| Mix2:1 | 0.26[1] | 0.59[2] | 1.4[3] |
| Mix1:1 | 0.23[1] | 0.43[2] | 1.4[3] |
| Nylon | 1.1[1] | 2.6[4] | 1.7[4] |
| Wood | 0.74[1] | 0.84[5] | 1.5[5] |

1: Values for the material of this study. 2: Calculated according to $C = \sqrt{V_p^2 - \frac{4}{3}V_s^2}$. 3: Assumed value. 4: Marsh (1980). 5: Values of birch wood (Marsh, 1980).

Table A-1 Impact conditions and the crater depth of basalt target.

| Shot # | Projectile | | | Depth |
|---|---|---|---|---|
| | Type[1] | m (g) | V (km s$^{-1}$) | (cm) |
| 201809-1 | g | 0.043 | 5.23 | 1.14 |
| 201809-2 | g | 0.043 | 5.13 | 1.02 |
| 201811-1 | g | 0.043 | 5.04 | 0.73 |
| 201811-2 | g | 0.043 | 5.03 | 0.78 |

1 g: glass spheres with diameter 3.2 mm.

Table A-2 Impact conditions and the largest-fragment mass fraction of pumice target.

| Shot # | Target | | Projectile | | | $M_1/M$ |
|---|---|---|---|---|---|---|
| | Size (cm) | M (g) | type[1] | m (g) | V (km s$^{-1}$) | |
| 060418-3 | 6.0×6.0×6.0 | 144 | n7 | 0.213 | 2.19 | 0.123 |
| 060418-4 | 6.0×6.0×6.0 | 147.8 | n7 | 0.213 | 2.58 | 0.135 |
| 060822-1 | 4.0×4.0×4.05 | 39.35 | n7 | 0.213 | 1.51 | 0.085 |
| 060822-2 | 4.0×4.0×3.95 | 41.65 | n7 | 0.213 | 2.13 | 0.037 |
| 060822-4 | 4.05×4.05×4.03 | 38.74 | g | 0.044 | 3.27 | 0.156 |
| 060824-5 | 4.05×4.05×4.05 | 40.19 | g | 0.044 | 1.64 | 0.983 |
| 060824-6 | 4.0×4.0×4.0 | 40.09 | g | 0.044 | 4.47 | 0.125 |
| 060825-4 | 4.03×4.03×4.03 | 38.7 | n7 | 0.213 | 3.28 | 0.028 |
| 70427 | 4.0×4.0×4.0 | 37.28 | n | 0.018 | 3.94 | 0.693 |
| 100602-1 | 4.0×4.0×4.0 | 42.28 | n7 | 0.219 | 7.10 | 0.007 |
| 100602-3 | 4.0×4.0×4.0 | 43.53 | n7 | 0.219 | 7.04 | 0.002 |

1 n7: nylon sphere with diameter 7 mm, g: glass sphere with diameter 3.2 mm, n: Nylon sphere with diameter 3.2 mm.

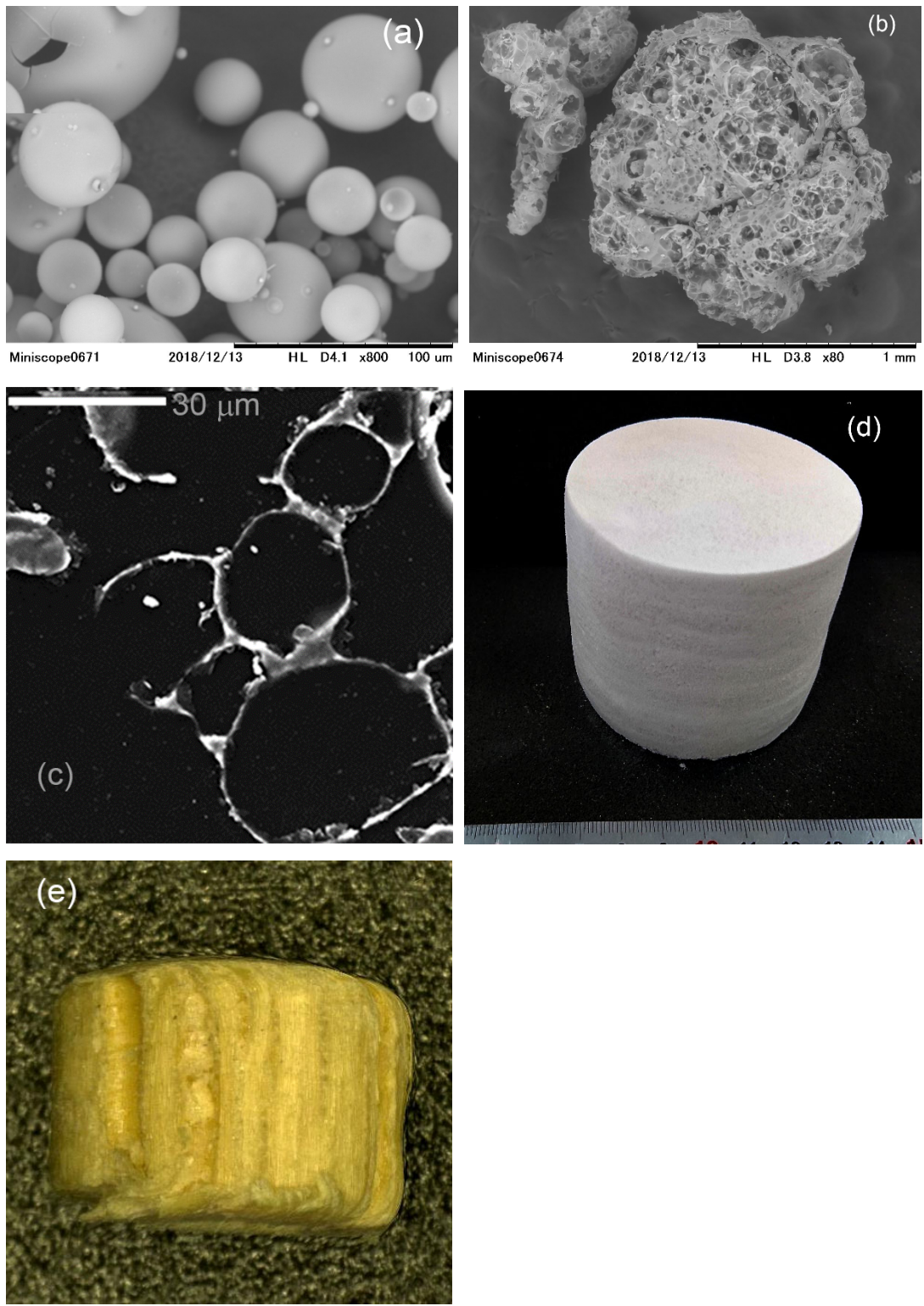

Fig. 1 Images of target and projectile. (a) SEM image of hollow glass beads. (b) SEM image of perlite grains. (c) SEM image of internal structure of HGB87. (d) Target (mix2:1). (e) Wood projectile.

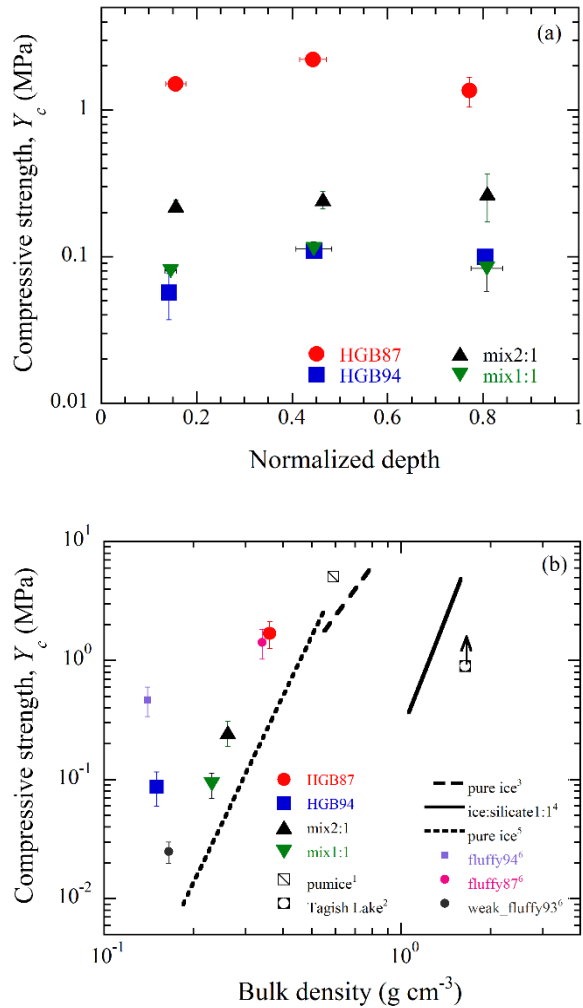

Fig. 2 (a) Compressive strength of the targets used in this study versus depth where the cylindrical specimens were drilled out. (b) Comparison of compressive strength and bulk density of the targets used in this study and other porous materials used in previous studies. Tagish Lake data is shown as a reference. 1 and 6: Okamoto and Nakamura (2017), 2: Hildebrand et al. (2006) for density and Tsuchiyama (private communication) for the lower limit of compressive strength (the original data was of crush strength), 3 and 4: Arakawa and Tomizuka (2004), 5: Shimaki and Arakawa (2012b).

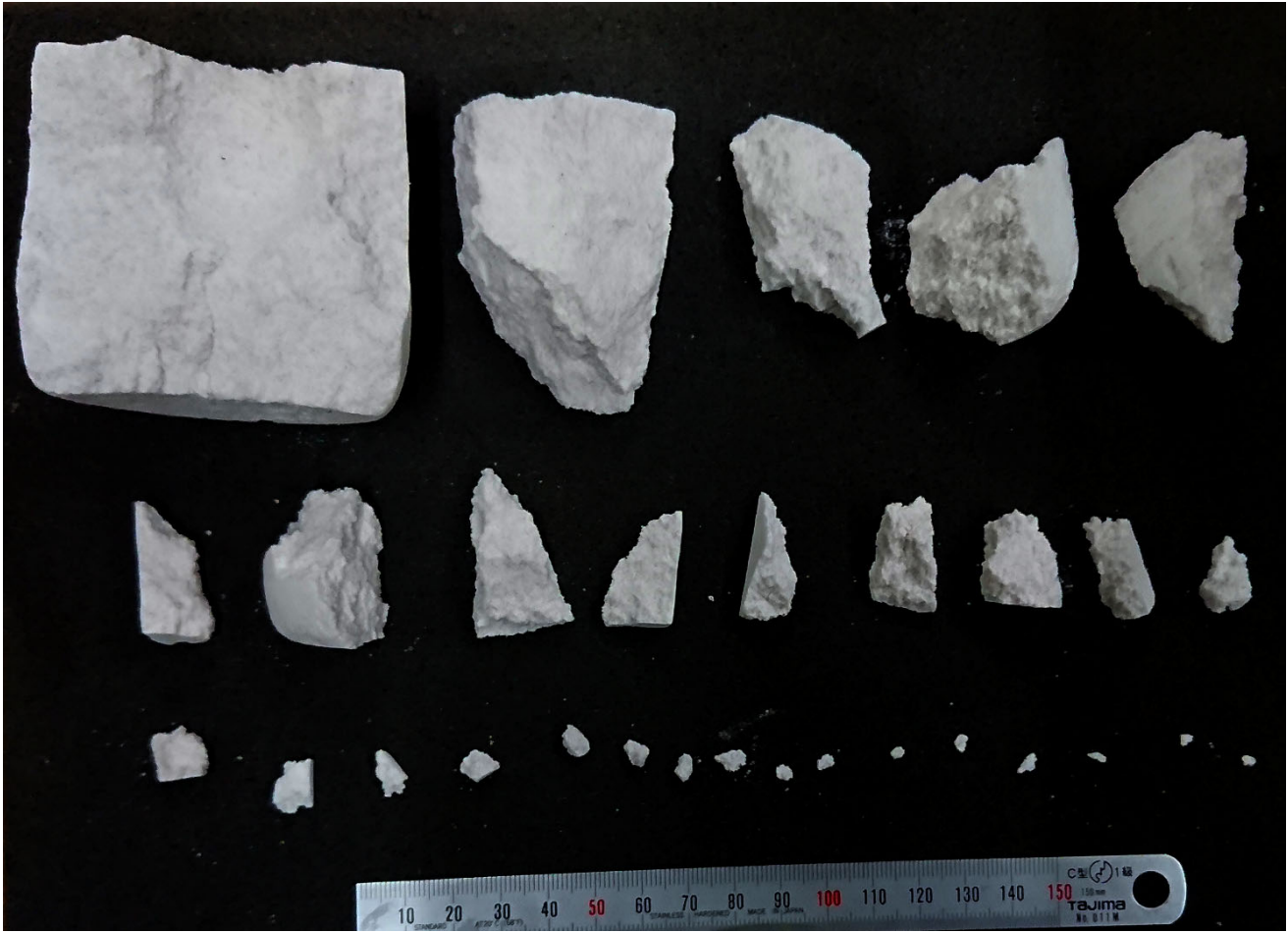

Fig.3 Fragments collected after impact (shot #m53). Bulb shaped cavity was formed in the largest fragment (top left corner).

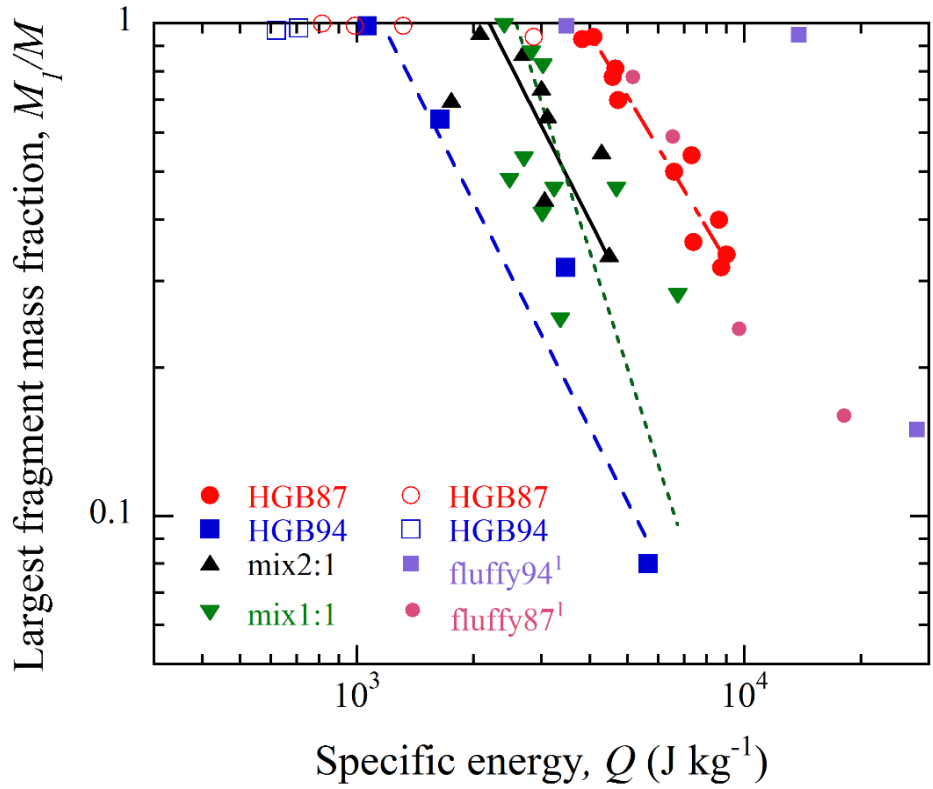

Fig. 4 Largest fragment mass fraction versus specific impact energy. Dash-dot, dashed, solid, and dotted lines show the least square fits to the data by Eq. (6). Filled marks show data used in the fitting, while open marks show those not included in the fitting. 1: Data of previous study of the targets with similar porosity to HGB94 (fluffy94 with axial ratio of the target of ~0.9) and the targets with similar porosity and strength to HGB87 (fluffy 87 with axial ratio of the target of ~0.8) (Okamoto et al., 2015).

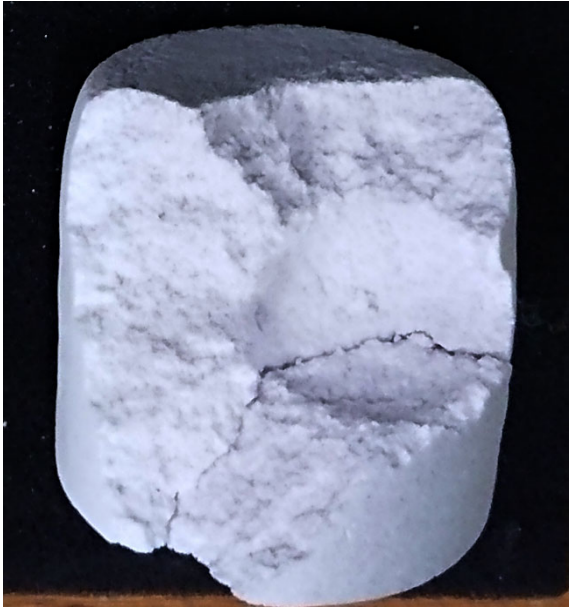 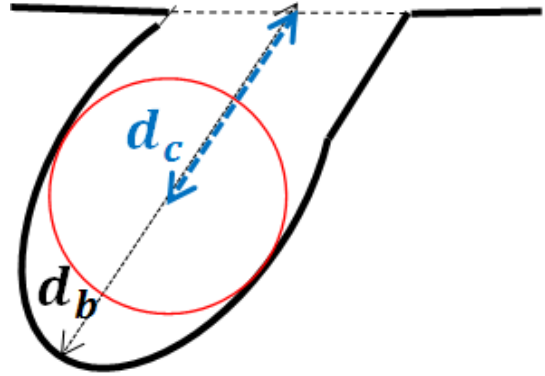

Fig. 5 (a) Spherical cavity formed in shot #m39. Projectile hit at the right side from lower than the horizontal direction. (b) Definition of cavity depth $d_b$ and the depth of the maximum diameter of cavity $d_c$.

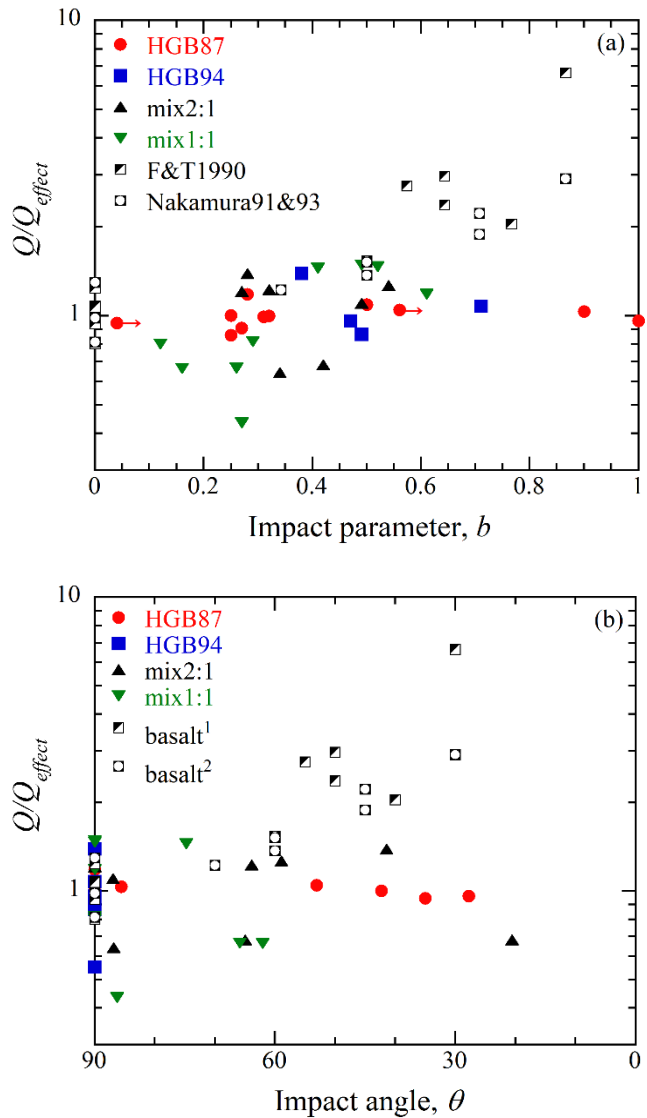

Fig. 6 Effect of oblique incidence. (a) Ratio of the actual specific energy to the one required to obtain the largest fragment $Q_{effect}$ versus impact parameter and (b) impact angle. Data indicated by filled marks in Fig. 4 are plotted here. Previous results of spherical basalt targets (1: Fujiwara and Tsukamoto, 1980; 2: Nakamura and Fujiwara 1991 and Nakamura 1993) are also shown. Empirical relationships $\frac{M_1}{M} = 1.96(\pm 0.14) \times 10^3 Q^{-1.23(\pm 0.05)}$ for basalt[1] and $\frac{M_1}{M} = 1.40(\pm 0.11) \times 10^3 Q^{-1.08(\pm 0.40)}$ for basalt[2] were derived based on the results of vertical shots of sphere targets of Fujiwara and Tsukamoto (1980) and Nakamura (1993), respectively, and used to calculate $Q_{effect}$.

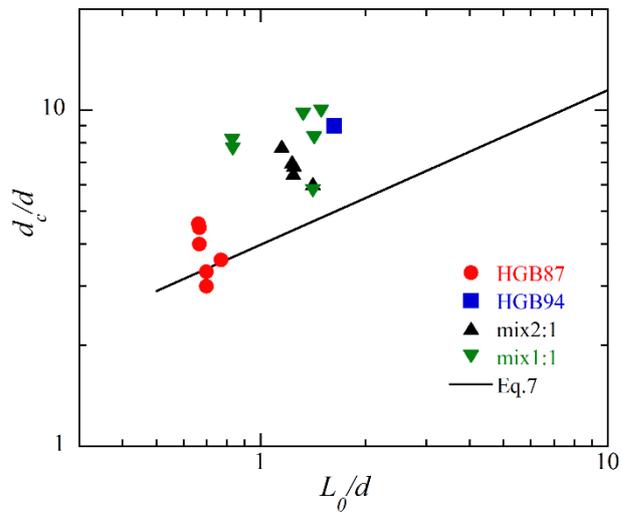

Fig. 7 Normalized depth at the maximum diameter of cavity $d_c/d$ versus normalized characteristic length. Solid line is the empirical relationship (Eq. 7) obtained in previous study (Okamoto et al., 2015).

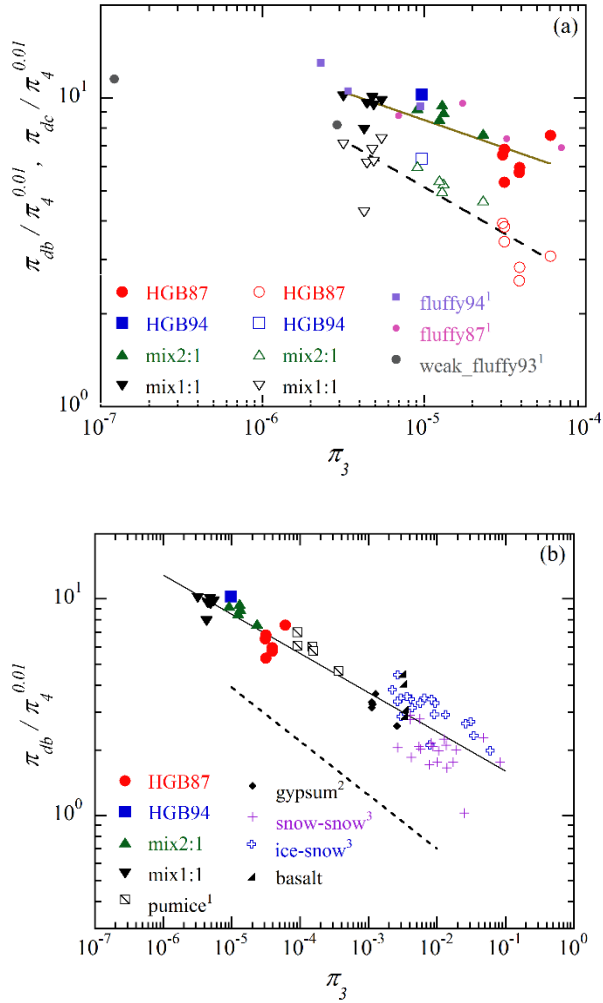

Fig. 8 (a) Normalized depth of cavity $\pi_{db}$ (filled marks) and normalized the depth at the maximum diameter of cavity $\pi_{dc}$ (open marks) versus $\pi_3$. Previous data (fluffy94, fluffy87, weak_fluffy93) (Okamoto and Nakamura, 2017) are also plotted. Solid line and dashed line are fitted curves to the data of this study. (b) Normalized depth of cavity for porous targets. 1: Okamoto and Nakamura (2017), 2: Okamoto and Nakamura (2017) and Kadono et al. (2018). Compressive strength of 15.6 MPa obtained in a previous study (Fujii and Nakamura, 2009) was assumed, although the target density is slightly different. 3: Arakawa and Yasui (2011). Strength based on dynamic measurement is adopted. Solid line is the same as in (a) and dotted line is the empirical relationship previously obtained for sedimentary rock, in which the tensile strength instead of compressive strength of target was used in $\pi_3$ (Suzuki et al., 2012). Experiment of basalt target is described in Appendix.

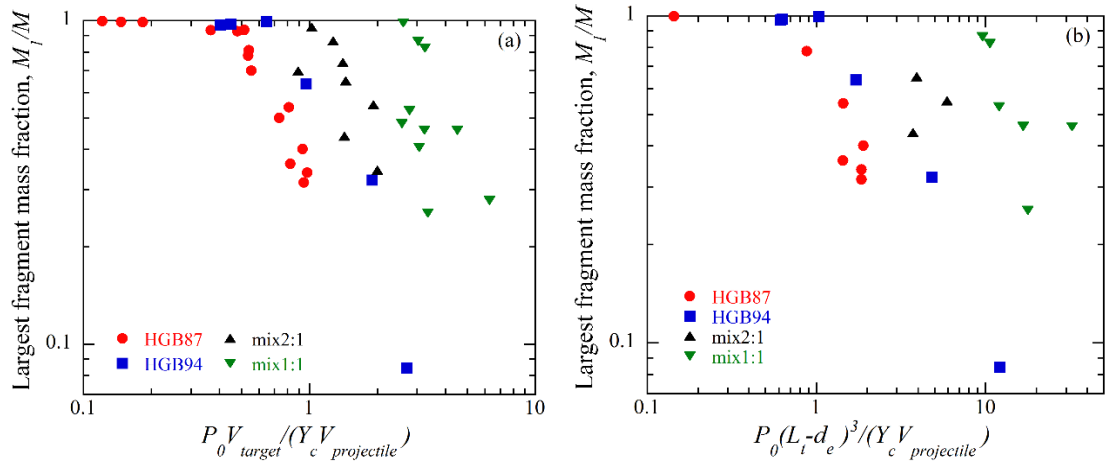

Fig. 9 Largest fragment mass fraction versus (a) non-dimensional impact stress, $P_{Ic}$, and (b) corrected version of non-dimensional impact stress where the depth of the center of the isobaric core is take into account.

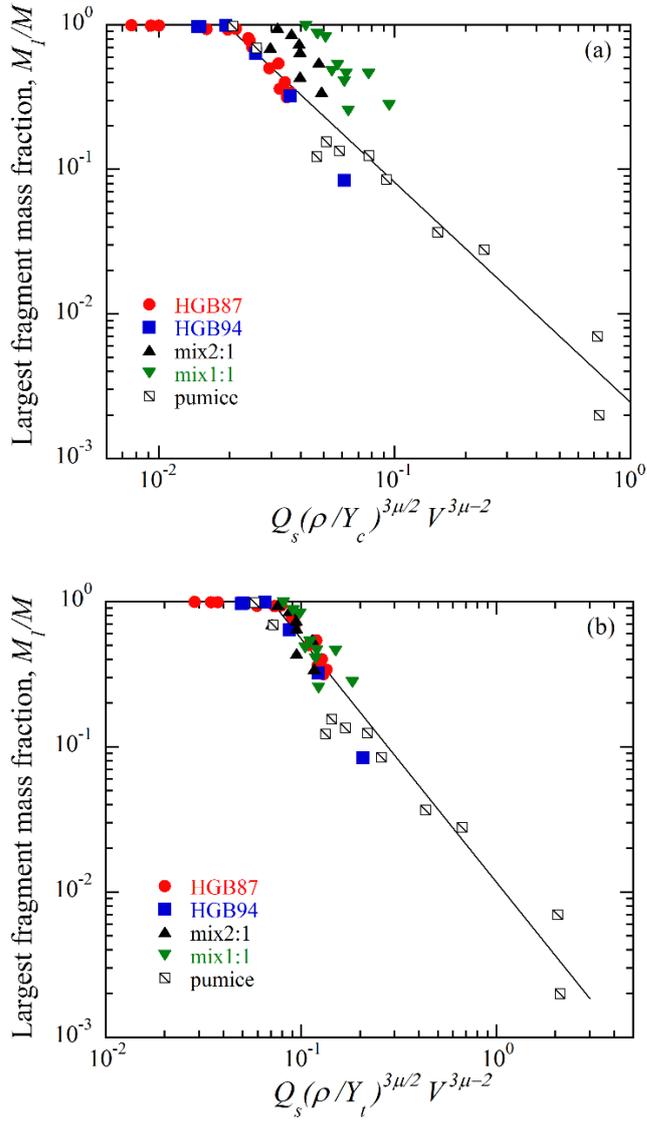

Fig. 10 Largest fragment mass fraction versus PI-group scaling parameters in which (a) compressive strength $Y_c$, and (b) tensile strength $Y_t$ is taken as the strength, respectively. The values of $\mu$ was assumed $\mu = \frac{2}{3} - \frac{1}{3}\phi$.